\documentclass{ws-p10x7}
\begin{document}

\def\P{{\rm I\kern-.15em P}}
\def\R{{\rm I\kern-.15em R}}
\def\q2{$Q^2$}
\newcommand {\xpom} {\mbox{$x_{_{\P}}$}}

\title{Diffractive phenomena}

\author{Giuseppe Iacobucci}

\address{INFN sezione di Bologna, Via Irnerio 46, I-40126 Bologna, Italy
\\E-mail: giuseppe.iacobucci@desy.de}

\twocolumn[\maketitle\abstract{
The most recent theoretical and experimental results in the field of 
diffractive scattering are reviewed.
%
A parallel between the two current theoretical approaches to diffraction, 
the DIS picture in the Breit frame and the dipole picture in the target frame,
is given, accompanied by a description of the 
models to which the data are compared.
A recent calculation of the rescattering corrections, 
which hints at the universality of the diffractive parton 
distribution functions, is presented.
The concept of generalized parton distributions is discussed
together with the first measurement of the processes 
which might give access to them.   
Particular emphasis is given to the HERA data, to motivate why
hard diffraction in deep inelastic scattering is viewed as an unrivalled
instrument to shed light on the still obscure aspects of hadronic 
interactions. 
}]

\section{Introduction}

Hadronic diffractive scattering was one of the most popular fields of
study in high-energy physics thirty years ago. 
The reason is probably that
Regge theory, at that time the leading theory of hadronic interactions,
was able to describe diffractive processes
in even more detail than more inclusive processes, such as the
total cross section $\sigma_{tot}$. 
The advent of quantum chromodynamics (QCD),
the quantum field theory of strong interactions, marked a new epoch,
since the perturbative QCD (pQCD) treatment reaped a harvest
of calculations and predictions in hadronic hard scattering 
that raised the strong interactions
to a level of understanding comparable to that reached by the
other interactions that form the Standard Model of particle physics.
However,  in high-energy scattering processes
QCD is applicable only when perturbative methods can be used,
i.e. in the presence of a {\it hard scale}.
This situation corresponds to small distance processes, or, equivalently,
large momenta involved ($p_T^2, Q^2, ...$).  
As a consequence, 
we are unable to use QCD to compute the bulk of hadronic interactions,
i.e. the {\it soft} --or large distance-- 
total, elastic and diffractive cross sections.
What happens is that at large distances the QCD coupling $\alpha_s$
becomes large and
the phenomenon of {\it confinement} of hadrons changes radically the
colour radiation pattern, making the calculations unaccessible.

Traditionally, confinement is studied by the investigation of
the binding forces between quarks, 
described in terms of interquark potentials,
which allow to calculate {\it static} properties of hadrons, such as the masses.
In high-energy hadronic scattering, there is a special class of events, 
that we call {\it hard diffraction}, 
in which a hard scale is present
and, at the same time, an initial state hadron may
emerge intact in the final state. For these events, 
the strong confinement forces prevail over the strong forces which tend to
break up the hadrons. By studying this class of events, 
we hope to learn about the fundamental properties of the binding forces
within hadrons.

In the following, mostly HERA data will be discussed, 
since the recent Tevatron data were 
already presented\cite{halina} at the Lepton-Photon conference two years ago.
In the coming Tevatron Run II, both the 
D0 and CDF detectors will be instrumented with forward proton spectrometers,
which will enhance their sensitivity to diffractive processes.

\subsection{Definition of diffractive scattering}\label{subsec:def}

\begin{figure}
\epsfxsize190pt
\figurebox{120pt}{190pt}{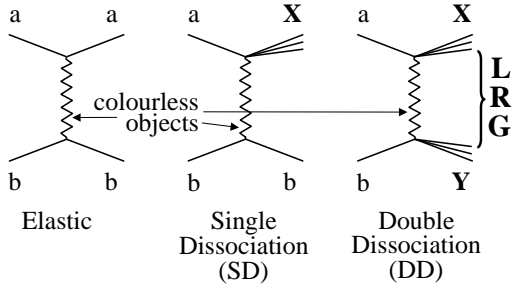}
\caption{Diagrams for the elastic (left), single dissociation (centre) and
double dissociation (right) reactions.}
\label{fig:diagrams}
\end{figure}

From the experimental point of view,
diffractive scattering can be defined as 
the sample of events ($\sim 30\%$ of the total cross section) in which:
{\it i)} the beam particles either remain intact (elastic) or dissociate 
into one or two hadronic systems (X, Y in Fig.~\ref{fig:diagrams}) 
of mass $M_{X,Y}^2$ much smaller than the centre-of-mass energy $s$;
{\it ii)} there is a $t$-channel colour-singlet exchange;
{\it iii)} the emerging systems hadronise independently, 
producing a large rapidity gap (LRG) in the distribution of the
final-state particles 
if the centre of mass energy $s$ is large enough 
(for single dissociation $y \approx \frac{1}{2} \cdot ln\frac{s}{M_X^2})$.
The experimental observations in diffractive hadronic scattering embrace 
a weak energy dependence of the cross section ($\sigma \propto s^{0.16}$),
very small scattering angles (parameterised by an exponential dependence 
of $t$, the square of the four-momentum exchanged, $e^{-b(s)\cdot |t|}$) 
and $t$-slopes, $b$, which are measured to depend on $ln(s)$ 
(often called shrinkage, since it is manifest as a shrinkage of 
the forward ($t=0$) elastic peak with increasing $s$).

\section{The hadronic language: \\
the Regge theory
and the parameters $\alpha_\P(0)$ and $\alpha^\prime_\P$}\label{sec:regge}

The experimental observations mentioned above are successfully 
described within the framework of the Regge theory. 
This is 
a theory of scattering in the complex angular momentum plane,
in which hadronic scattering
is described by the exchange of collective states,
called trajectories, which are made by families of particles 
whose spins are measured to be in linear relation 
with their squared masses at small values of $|t|$.
This property allows one to parameterise the trajectories as
$\alpha_j(t) = \alpha_j(0)+\alpha^\prime_j \cdot t$,
where $j$ can be a pion, a pomeron ($\P$) or a reggeon ($\R$).
As an example, 
the cross section for the elastic process $ab \rightarrow ab$ 
can be written, for small $|t|$, in Regge theory as
$\frac{d\sigma_{el}^{ab}}{d|t|} = \frac{1}{16\pi}\cdot \sum_{j}[\beta_{aj}(0) \cdot \beta_{bj}(0)]^2 \cdot s^{2(\alpha_j(0)-1)} \cdot e^{-b(s) \cdot |t|}$, 
where the (``residue'') functions $\beta(t)$ represent the couplings,
$b(s) = b_a + b_b +  2\alpha^\prime_j \cdot ln(s)$
and $b_a$ and $b_b$ originate from the form factors of the hadrons $a$ and $b$.
The formula for $\frac{d\sigma_{el}^{ab}}{dt}$ describes 
the $s$ and $t$ dependence noted in Sect.~\ref{subsec:def}

\begin{figure}
\epsfxsize190pt
\figurebox{120pt}{190pt}{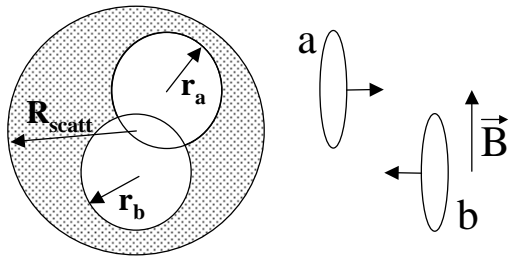}
\caption{Cartoon of hadronic scattering in the transverse (left) and
longitudinal (right) plane.}
\label{fig:cartoon}
\end{figure}

The $\P$ trajectory\footnote{The $\P$ trajectory carries quantum numbers
C = P = +1. When it was introduced, it was 
the only one not associated with any real particle.
Since then, the effort to find particles associated to the 
$\P$ trajectory, the so called {\it glueballs}, has been continuous. 
Today's situation is reviewed in
ref.~\protect\cite{close}.}
was postulated with an intercept $\alpha_\P(0)$ larger than 1
to fit the hadronic total cross sections which increase at high energy.
The simultaneous use of the $\P$ and $\R$ trajectories fitted 
the $s$ dependence of the
total cross section in $pp, \pi p, Kp$ and $ \gamma p$ scattering
using the simple form
$\sigma_{tot} = X\cdot s^{\alpha_\P(0)-1} + Y\cdot s^{\alpha_\R(0)-1}$,
where $X$ and $Y$ are constants.
The values  $\alpha_\P(0) = 1.08$ and $\alpha^\prime_\P=0.25$
obtained by the fits of Donnachie and Landshoff~\cite{DL},
are often referred to as the {\it soft} $\P$.
The fact that the high-energy behaviour of all the measured hadronic total 
cross sections could be described by the same $\alpha_\P(0)$
is a great success for Regge phenomenology and
shows that these two parameters are not just incidental
parameters obtained by a phenomenological fit, but they have a deeper meaning. 
Indeed, they must be fundamental,
since they represent universal (i.e. independent of the initial state hadrons)
features of the strong forces that govern the binding 
of hadrons\cite{bartels}.
The {\it intercept} 
$\alpha_\P(0) = 1+\epsilon$
governs the energy dependence of 
the total $(\sigma_{tot}\propto s^\epsilon)$, 
elastic and diffractive 
$(\sigma_{el}$, $\sigma_{diffr} \propto s^{2\epsilon})$ cross sections.
Therefore, in a geometrical interpretation,
it can be related
to the transverse extension of the scattering system
(${\rm R_{scatt}}$ in Fig.~\ref{fig:cartoon} left),
i.e. of the colour-radiation cloud, which increases with energy.
%
The {\it slope} $\alpha^\prime_\P$ on the other hand determines the
dependence of the cross section 
on the impact parameter $\vec B$ (Fig.~\ref{fig:cartoon} right), 
$\langle\vec B^2\rangle = 2 b(s) =
2 [b_a + b_b + 2 \alpha^\prime_\P \cdot ln(s)]$ as described in~\cite{bartels}.
As such, $\alpha^\prime_\P$ reflects the strength of the binding forces
and characterizes the confinement forces in QCD.
It is important to stress here that only diffractive interactions give
access $\alpha^\prime_\P$, since its measurement requires semi-inclusive 
diffractive processes in order to measure $t$.

The calculation of $\alpha_\P(0)$ and $\alpha^\prime_\P$ 
from first principles is of primary importance. 
Many efforts have been going on in recent years, 
based on novel physical concepts
and mathematical methods.
As an example, Kharzeev et al.\cite{kharzeev} introduced a new type of 
instanton-induced interaction (instanton ladder) which is 
proposed to be responsible for the structure of the soft $\P$.
The work by Witten\cite{witten}, in which confinement and mass gap 
(but not asymptotic freedom and mass spectra)
are obtained in the framework of string theory, 
shows how diverse and basic are the attempts to go beyond pQCD.

\section{The partonic language: \\ 
hard diffraction and QCD}\label{subsec:qcd}

So far, we discussed soft-hadronic interactions,
which are governed by the hadronic degrees of freedom.
However, to understand the dynamics,
we need to describe hadronic phenomena in terms
of hadron sub-components and quantum-field theories, 
i.e. in terms of partons and QCD. 
Since we are able to apply perturbative methods to QCD only 
if a hard scale is present,
the theory is currently limited to the calculation of hard interactions.
As we will see in this Section, there are two approaches to
diffraction in the presence of a hard scale, 
which have their interpretation in different reference frames.
But, before discussing the theory of hard diffraction, 
let us review the advantages 
that HERA offers to this field.

\subsection{Diffractive scattering at HERA}\label{subsec:hera}

The HERA collider at DESY, born to study the proton structure functions
and to search for exotic processes,
acquired a central r\^ole in the attempt to understand diffractive
interactions. 
There are several advantages in studying diffractive hard scattering at HERA.
Deep-inelastic scattering (DIS) is much simpler than hadron-hadron collisions,
since only one hadron, i.e. one large ($\sim 1$~ fm$)$
~non-perturbative object, is present in the initial state.
The huge range in resolution power,
$0 < Q^2 < 10^{5}$~ GeV$^2$
($Q^2$ is the negative squared four
momentum of the exchanged virtual photon, $\gamma^\star$),
which corresponds to probing distances of $10^{-3} < \Delta r <1$~fm,
allows the investigation of  both the short and long distance regions.
The small-$x$ values reachable at HERA, 
where $x$ is the Bjorken DIS scaling variable, 
correspond to a region of high
parton densities; thus information on the saturation 
(and possibly confinement)
of partons in the proton might be accessed.
The asymmetric beam energies 
($E_e = 27.5$~GeV and $E_p = 820$ or $920$~GeV)
offer an excellent experimental acceptance for the 
$\gamma$-diffractively-dissociated system,
thus opening up the photon hemisphere.
Finally, about $10\%$ of the DIS events at small-$x$ at HERA are diffractive,
providing large samples of events to study.

It is important to mention that the universality of 
$\alpha_\P(0)$ and $\alpha^\prime_\P$ experimentally holds as long as we 
consider the scattering between hadrons. 
At HERA, where a virtual photon emitted by the beam lepton 
scatters off a proton,
the measured $\P$ trajectory\footnote{Even if it is still possible to use the
$\P$ language in diffractive hard scattering,
there are models which do not make 
use of the concept of $\P$. In the following, I will continue to use the
term ``$\P$'', but the reader should remember that this will not be the
soft $\P$ we discussed in Section~\ref{sec:regge}.} 
is found to depend on the photon 
virtuality\cite{arik} or on the diffractive final 
state\cite{rhotraj,jpsitraj}. 
This result is expected\cite{bartels} to provide information, 
otherwise inaccessible,
on the dynamics of strong interactions.

\subsection{The diffractive structure functions}\label{subsec:dsf}

\begin{figure}
\epsfxsize120pt
\figurebox{120pt}{190pt}{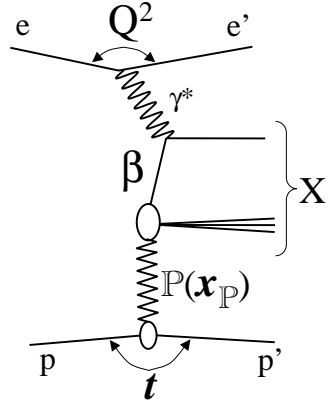}
\caption{Schematic view of diffractive deep inelastic $ep$ scattering.}
\label{fig:diffdis}
\end{figure}

The cross section for the diffractive process $ep \rightarrow eXp$ 
in DIS can be written: \\
\begin{eqnarray}
&&\frac{d^4\sigma_{diffr}}{d\beta dQ^2 d\xpom dt} = \nonumber \\[4pt]
&&~~\frac{2 \pi \alpha^2}{\beta Q^4} \; (1+(1-y)^2) \;
F_2^{D(4)}(\beta,Q^2,\xpom,t),\label{eq:f2d4}
\nonumber 
\end{eqnarray} 
\noindent 
\noindent
where $\alpha$ is the electromagnetic coupling constant, 
$y$ the inelasticity and 
the contributions of the longitudinal proton structure function and of
$Z^0$ exchange have been neglected.
The diffractive structure function $F_2^{D(4)}$ depends on four variables.
As can be seen in Fig.~\ref{fig:diffdis},
in first approximation,
\q2 and $\beta$ describe the photon vertex 
while \xpom~and $t$ describe the proton vertex.
In a reference frame in which the proton is fast, 
Breit or Bjorken-DIS frames ({\it DIS picture}),
$\beta = \frac{Q^2}{M_X^2+Q^2}$
can be identified with the fraction of the $\P$ 
momentum carried by 
the quark which couples to the $\gamma^\star$
and $x_{_\P} = \frac{M_X^2+Q^2}{W^2+Q^2}$ 
with the fraction of the proton 
momentum carried
by the $\P$.
In the previous formula, $W$ is the $\gamma p$ centre of mass energy.
Integration over $t$ gives $F_2^{D(3)}(\beta,Q^2,\xpom)$.

\subsection{Models for hard diffraction}\label{subsec:models}

The era of hard diffraction started in 1987, 
when the UA8 collaboration at the SPSC measured diffractive jet production
in $p\bar p$ scattering\cite{ua8}. This process was predicted by 
Ingelman and Schlein\cite{IS} a few years earlier,
in the framework of a model in which  the concept of 
the soft $\P$ was applied to hard diffractive scattering.
The Ingelman-Schlein model was based on several assumptions. 
It was assumed that 
the soft $\P$ has a partonic structure 
like a real hadron, that Regge factorisation holds, 
i.e. that the diffractive structure function can be factorised into
$F_2^{D(4)}(\beta,Q^2,\xpom,t) = \Phi_{\P/p}(x_{_\P},t)\cdot 
F_2^{D(2)}(\beta,Q^2)$, and that the $\P$ flux is 
the one given by Regge theory,
$\Phi_{\P/p}(x_{_\P},t) = (1/x_{_\P})^{2\alpha_{_\P}(t)-1}$.

In recent years there has been a lot of progress in the theory of
diffractive DIS.
Hard QCD factorisation:
\begin{eqnarray}
&&\frac{d^4\sigma_{diffr}}{d\beta dQ^2 d\xpom dt} = \nonumber \\[4pt]
&&\sum_{i} \int_x^{x_{_\P}} dx^\prime 
\hat \sigma_{hard} (x,Q^2,x^\prime) \cdot 
\frac{df_i^D(x^\prime,x_{_\P},t)}{dx_{_\P} dt}\nonumber 
\end{eqnarray} 
\noindent 
was proven\cite{collins} for large enough \q2.
Here 
$\hat \sigma_{hard}$
is the partonic hard cross section.
The non-perturbative diffractive parton distributions\cite{berera} (DPD's)
$\frac{df_i^D(x^\prime,x_{_\P},t)}{dx_{_\P} dt}$
are conditional probabilities of finding, in a fast proton,
a parton $i$ with longitudinal momentum fraction $x^\prime$, 
while at the same time the beam proton is scattered with $t$ and $(1-x_{_\P})$.
DPD's were also found to obey the usual 
DGLAP evolution equations\cite{collins}.
These results allow the assertion that,
like inclusive DIS, diffractive DIS is firmly rooted 
in QCD.

\begin{figure}
\epsfxsize190pt
\figurebox{120pt}{190pt}{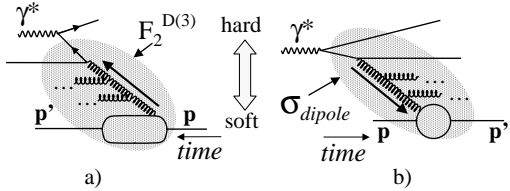}
\caption{Comparison between the DIS (a) and the dipole (b) pictures.}
\label{fig:compare}
\end{figure}

In a frame where the proton is fast,
diffractive DIS can be described as follows (see Fig.~\ref{fig:compare}a).
A parton is emitted by the fast proton  
and then evolves into a parton cascade 
which eventually 
produces a $q \bar q$ pair that interacts with the $\gamma^\star$.
The parton emission in this frame is ordered in time 
from the proton to the $\gamma^\star$ side.
The presence of a hard scale solves only part of the problem
by making the upper part of the diagram 
in Fig.~\ref{fig:compare}a 
calculable in pQCD,
while the lower, soft part of the interaction, 
parameterized by the DPD's, occurs over a 
large space-time and a perturbative treatment is not possible.

The physical picture of diffractive DIS 
is most easily seen in models built in the
proton rest frame (Fig.~\ref{fig:compare}b). 
In this frame, the $\gamma^\star$ fluctuates into 
a $q \bar q$ ($ q \bar q g, ...$) system of lifetime
$\tau \approx 1/x$. At the small-$x$ values available at HERA, 
this lifetime corresponds to a large distance, up to 1000 fm. 
Thus, from the point of view of the proton, the $q \bar q$ system is frozen
and the interaction happens between a colour dipole and the proton.
As a consequence, these models are named {\it dipole models}.
The diffractive cross section in the target frame can be written:
\begin{eqnarray}
\sigma = 
\int d^2r~dz \cdot
|\psi(r,z,Q^2)|^2 \cdot \sigma_{dipole}^2(x,r) \nonumber
\end{eqnarray} 
\noindent 
where $\psi$ is the $\gamma^\star \rightarrow q \bar q$ wave function,
$z$ is the fraction of the $\gamma^\star$ momentum carried 
by one of the quarks 
and $r$ is the dipole size, 
i.e. the transverse separation of the $q \bar q$ system.
The non-perturbative structure is contained in 
the dipole cross section, $\sigma_{dipole}$, which is modelled at the 
lowest order in pQCD by a two-gluon exchange.
The time ordering of the parton emissions in the target frame is 
reversed with respect to the DIS frame: 
in the target frame, the $\gamma^\star$ fluctuates into a
$q \bar q$ colour dipole which develops a
parton cascade and finally interacts with the proton.

It should be noted that, 
despite the apparent differences between the two approaches to DIS, 
the physics must be the same. 
A theoretical effort\cite{collins2} is indeed going on, to establish a
correspondence between NLO and large $r^2$ in the two approaches.

\section{Structure functions and jet measurements in diffractive DIS}

Three different methods are used to identify diffractive events at HERA:
{\it i)} the scattered proton is measured in spectrometers
positioned along the proton beamline, 
such as the ZEUS LPS\cite{lps} and the H1 FPS\cite{fps};
{\it ii)} a LRG in the final state is required in the direction of the outgoing
proton. This is achieved\cite{lrg} by requiring 
no energy deposits above a certain threshold (e.g. $400$~MeV) 
at pseudorapidities smaller than a value $\eta_{max}$, 
thus imposing no-hadronic activity around the scattered proton;
{\it iii)} the diffractive sample is separated from 
the non-diffractive events by using the fact 
that the non-diffractive sample has an exponentially falling dependence on 
$M_X^2$, while the diffractive part has
an approximate $1/M_X^2$ dependence\cite{mx2}.

\subsection{$F_2^D$ measurements}

\begin{figure}
\epsfxsize200pt
\epsfysize230pt
\figurebox{260pt}{190pt}{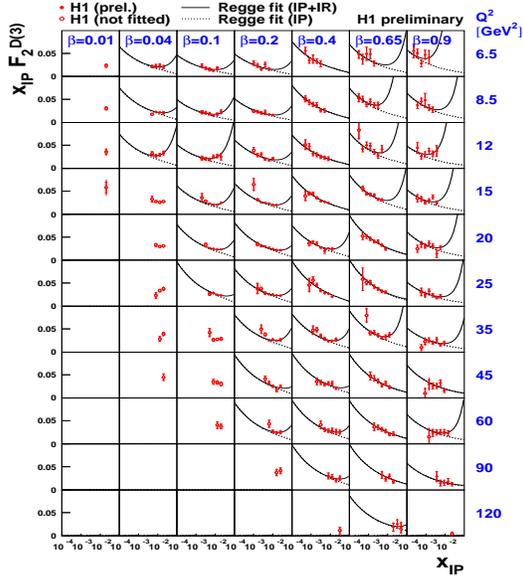}
\caption{$x_{_\P}F_2^{D(3)}$ vs. 
$x_{_\P}$ at fixed values of $\beta$ and $Q^2$.
The lines show the results of the fit described in the text
(full lines: $\P + \R$, dotted lines: $\P$ only). 
Open symbols indicate the points excluded from the fit.
}
\label{fig:h1f2dall}
\end{figure}

A new, high-precision measurement\cite{h1incl} of the diffractive~structure 
function $F_2^{D(3)}$
was performed by the H1 Collaboration in the kinematic range
$6.5 < Q^2 < 120~{\rm GeV}^2,~0.01 < \beta < 0.9, 
~10^{-4} < x_{_\P} < 0.05$ and $|t| < 1$ GeV$^2$.
The event selection requires a LRG in the final state,
thus selecting $ep \rightarrow eXY$ events,
where $Y$ is either a proton or a low-mass proton-excitation system.
The $x_{_\P}$ dependence of the data is well described
by Regge phenomenology (see Fig.~\ref{fig:h1f2dall}) if a leading ($\P$) 
and a secondary ($\R$) trajectory exchanges 
are assumed.
The effective pomeron intercept is
measured to be $\alpha_{_\P}(0) = 1.173 \pm 0.018(stat.) 
\pm 0.017(syst.) ^{+0.063}_{-0.035}(model)$.
The ratio 
$\sigma_{diffr}/\sigma_{tot}$
is relatively flat as a function of $W$.
The data are
well described by a fit based on the
DGLAP evolution of the $\beta$ and $Q^2$ dependence and
Regge factorisation, with 
a Regge motivated $(1/x_{_\P})^{2\alpha_{_\P}(t)-1}$ dependence.

\begin{figure}
\epsfxsize120pt
\figurebox{120pt}{190pt}{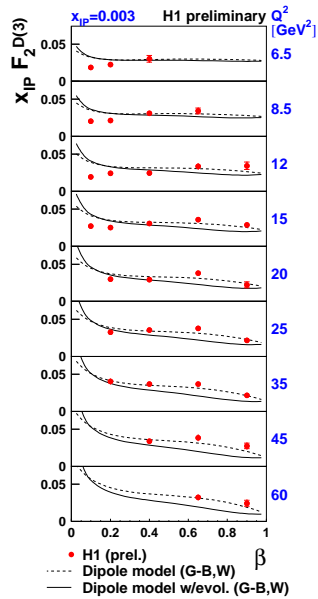}
\caption{$x_{_\P}F_2^{D(3)}$ at fixed $x_{_\P} = 0.03$ 
as a function of $\beta$ for various values of $Q^2$.
The lines show the prediction of the saturation model
without\protect\cite{gbw} (dashed lines) and with\protect\cite{gbw2} 
(full lines) QCD DGLAP evolution.}
\label{fig:h1f2dgbw}
\end{figure}
In Fig.~\ref{fig:h1f2dgbw}, the H1 data are compared with two versions of 
the {\it ``saturation''} model by Golec-Biernat and 
${\rm W\ddot usthoff}$\cite{gbw,gbw2}.
In this model, the unitarity  of $\sigma_{dipole}$ at small $x$,
that is the fact that the cross sections should not diverge 
at asymptotic energies as it happens in the QCD description
 based on linear evolution equations,
is imposed.
Unitarity 
is built by postulating the phenomenological form 
\begin{center}
$\sigma_{dipole}=\sigma_0\cdot[1-exp({\frac{-r^2}{4R_0(x)^2}})]$,\\
\end{center}
where $R_0(x)=\frac{1}{Q_0}\cdot(\frac{x}{x_0})^{\frac{\lambda}{2}}$
can be interpreted as the saturation radius, $Q_0 = 1$ GeV,
and $\sigma_0,~x_0$ and $\lambda$ are free parameters of the model.
At small $r$, the dipole cross section exhibits colour transparency
($\sigma_{dipole}\approx r^2$),
which is a purely pQCD phenomenon, while saturation
($\sigma_{dipole}\simeq \sigma_0$) occurs at large $r$.
The transition between the two regimes is governed by the
saturation radius $R_0(x)$.
The saturation model contains 
a higher-twist contribution at large $\beta$, which allows comparisons
to be made throughout the full measured kinematic region.
The values of the three parameters 
$\sigma_0,~x_0$ and $\lambda$ obtained 
by a fit to inclusive DIS data with $x<0.01$, 
were used to predict $F_2^D$.
The result, shown in Fig.~\ref{fig:h1f2dgbw}, 
gives a good description of the data
except at small $\beta$ and $Q^2$.
The model in which QCD DGLAP evolution is added\cite{gbw2}
(full lines in Fig.~\ref{fig:h1f2dgbw}) 
underestimates the measured $F_2^{D(3)}$ at high $\beta$ and $Q^2$.

The H1 data were also
compared with the {\it ``semi-classical''} model 
by ${\rm Buchm \ddot u ller}$ et al.\cite{semicl}.
In this model, the proton is seen as a superposition of soft colour
fields parameterised according to a simple non-perturbative model that 
averages over all field configurations.
In the target frame, 
the $q \bar q$ or $q \bar q g$ fluctuations of the $\gamma^\star$,
modelled as colour dipoles,
traverse the proton colour field. Diffractive (non-diffractive)
interactions occur if both colour dipole and target emerge in a 
colour singlet (octet) state\footnote{This model is particularly 
interesting since it made the prediction 
$\sigma_{diffr}/\sigma_{tot} \approx$ independent of $W$, 
which is found in the data\protect\cite{mx2,h1incl}.}.
The model, which  contains only four parameters obtained by a combined
fit to the $F_2$ and $F_2^D$ data,
reproduces the general features of the data\cite{h1incl} 
(not shown) but lies above
the data when $\beta$ and $Q^2$ are both small.

The process $ep \rightarrow eXp$
has been studied\cite{lpsnew} by the ZEUS Collaboration
selecting diffractive events by the requirement of a scattered proton 
track, carrying a fraction of the initial proton momentum $x_L > 0.95$,
in the ZEUS leading proton spectrometer (LPS).
\begin{figure}
\epsfxsize190pt
\figurebox{120pt}{190pt}{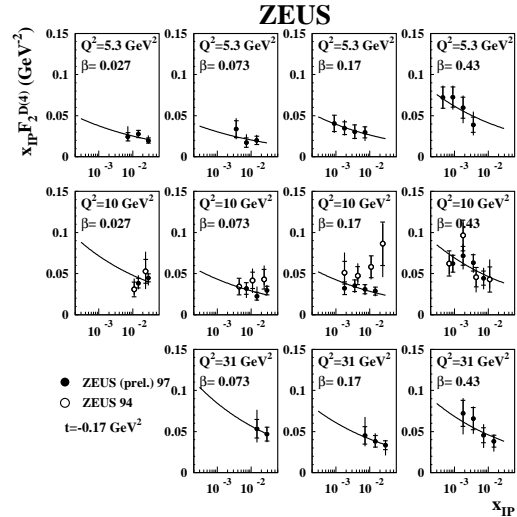}
\caption{ZEUS LPS measurement of $x_{_\P}F_2^{D(4)}$ (solid points)
as a function of $x_{_\P}$ for the $\beta, Q^2$ 
and $t$ values indicated in the plot.
Solid lines show the result of the fit described in~\protect\cite{lpsnew}.
Open points show the results obtained in the previous 
ZEUS LPS measurement\protect\cite{lpsold}.
}
\label{fig:zf2dlps}
\end{figure}
The detection of the final state proton in the LPS
(though substantially reduced the event sample 
because of the small LPS acceptance)
allows a direct measurement of $t$. It also provides the
cleanest selection of diffractive events, independent of the hadronic
final state and free of the proton-dissociation background events,
$ep \rightarrow eXY$.
The diffractive structure functions $F_2^{D(3)}$ and $F_2^{D(4)}$
were measured in the kinematic range
$4 < Q^2 < 100~{\rm GeV}^2,~0.01 < \beta < 0.6, 
~10^{-4} < x_{_\P} < 0.04$ and $0.075 < |t| < 0.035$ GeV$^2$.
Figure~\ref{fig:zf2dlps} shows the measured 
$x_{_\P}F_2^{D(4)}(\beta,Q^2,x_{_\P},t)$
as a function of $x_{_\P}$ for different values of $\beta$ and $Q^2$
and for average $t = -0.17$~GeV$^2$.
The $x_{_\P}$ dependence of the $F_2^{D(3)}$ obtained with the LPS
can be fitted using a flux factor $(1/x_{_\P})^{2\alpha_{_\P}(t)-1}$
(solid lines in Fig.~\ref{fig:zf2dlps}),
therefore showing consistency with Regge factorisation.
The value of the pomeron intercept obtained by this fit is
$\alpha_{_\P}(0) = 1.13 \pm 0.03(stat.)^{+0.03}_{-0.01}(syst.)$.

\subsection{Extraction of DPD's from inclusive diffraction}

H1 extracted the DPD's from inclusive diffractive data\cite{h1fit}.
\begin{figure}
\epsfxsize190pt
\figurebox{120pt}{190pt}{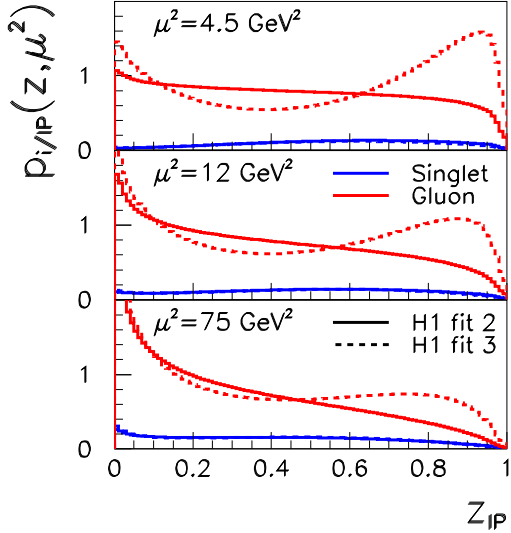}
\caption{The quark and gluon  
distributions extracted at various values of $Q^2$ for two different 
parameterisations of the parton densities.
}
\label{fig:gluon}
\end{figure}
They used a QCD-motivated model in which parton distributions,
which evolve according to the NLO DGLAP evolution equations, are
assigned to the leading ($\P$) and subleading ($\R$) exchanges
utilized to parameterise the data.
Under such a hypothesis, the data require approximately $90\%$ and $80\%$
of the momentum of the pomeron to be carried by gluons at $Q^2 = 4.5$~GeV$^2$
and $Q^2 = 75$~GeV$^2$, respectively (see Fig.~\ref{fig:gluon}).
The inclusive measurements are not particularly sensitive
to the shape of the gluon distribution at large $z_{_\P}$, the 
momentum fraction of partons in the 
pomeron\footnote{Notice that $z_{_\P} \equiv \beta$
if the parton in the $\P$ is a quark.}, and both the ``flat gluon'' 
(full lines in Fig.~\ref{fig:gluon})  and the ``peaked gluon'' 
(dashed lines in Fig.~\ref{fig:gluon}) fits give a good $\chi^2$.

\subsection{Diffractive jet production}

Diffractive dijet production shows higher sensitivity to the gluon
density in the $\P$ than the inclusive measurements.
The dijet cross sections have been measured\cite{h1jets} by H1 
in the kinematic range
$4 < Q^2 < 80~{\rm GeV}^2,~90<W<260~{\rm GeV},~23<M_X<40~{\rm GeV}$
and $x_{_\P} < 0.05$
for jets of $p_T>4~{\rm GeV}$ reconstructed using the   
cone algorithm ($R=1$) in the $\gamma^\star p$ reference frame.
\begin{figure}
\epsfxsize190pt
\figurebox{120pt}{190pt}{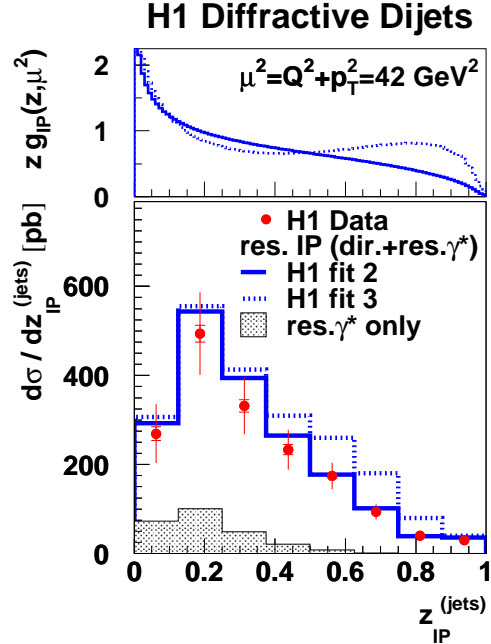}
\caption{The diffractive dijet cross section as a function of $z_{_\P}$.
The data are compared to the 
results obtained by the RAPGAP Monte Carlo (MC) using the
``flat gluon'' (solid line) and the ``peaked gluon'' (dotted line)
from the H1 QCD fit\protect\cite{h1fit}.
The corresponding gluon distributions are shown on the top plot.
}
\label{fig:jets}
\end{figure}
The H1 cross sections, compared in Fig.~\ref{fig:jets} 
to the RAPGAP\cite{rapgap} MC,
which is an implementation of the Ingelman-Schlein model,
favour the flat-gluon DPD's.
The dijet cross sections in bins of $Q^2+p_T^2$ and $x_{_\P}$ 
show\cite{h1jets} that the  DGLAP evolution holds 
and that the data are consistent with
Regge factorisation, respectively, as assumed in the RAPGAP MC.

\begin{figure}
\epsfxsize190pt
\figurebox{120pt}{190pt}{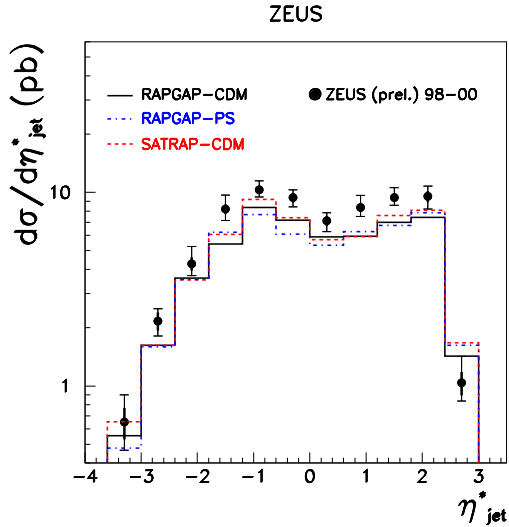}
\caption{The diffractive cross section for three jet events 
as a function of the pseudorapidity in the $\gamma^\star \P$ frame
of each jet, $\eta^\star_{jet}$.
}
\label{fig:3jets}
\end{figure}
ZEUS measured\cite{zeusjets} diffractive three-jet cross sections in the 
kinematic range
$5 < Q^2 < 100~{\rm GeV}^2,~200<W<~250~{\rm GeV}$ and 
$23<M_X<40~{\rm GeV}$.
The jets were reconstructed using the $k_T$ 
algorithm in the $\gamma^\star \P$ reference frame.
The result, shown in Fig.~\ref{fig:3jets},
is broadly consistent with models in which the hadronic final-state is
dominated by a $q \bar q g$ system with the gluon preferentially emitted
in the pomeron direction. Such configurations are predicted both by RAPGAP
and by the SATRAP\cite{satrap} MC, which is based on the saturation model.
The two MC models describe the data
equally well, except for a ($\approx 20\%$) normalisation factor, 
which can probably be ascribed to higher order corrections 
that are only included approximately by parton showers.

\section{Universality of the DPD's?}

\begin{figure}
\epsfxsize180pt
\figurebox{120pt}{190pt}{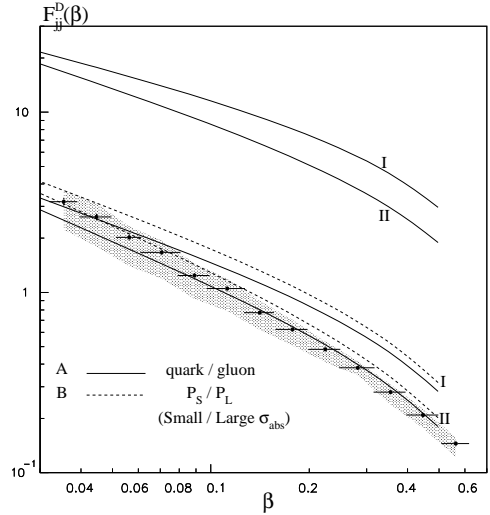}
\caption{The CDF measurement of $F_{jj}^D(\beta)$ (black circles)
compared with predictions\protect\cite{kaidalov}
obtained for two sets of HERA DPD's (I and II).
The upper two curves correspond to the neglect of rescattering corrections,
whereas the lower four curves show the effect of including these corrections
using two models (A and B) for the diffractive eigenstates.
}
\label{fig:cdf}
\end{figure}

The CDF Collaboration measured the diffractive structure functions from
dijet events, $F_{jj}^D(\beta)$, 
in $p \bar p$ collisions at the Tevatron.
The results\cite{cdf}, shown in Fig.~\ref{fig:cdf}, 
are a factor of ten smaller than the predictions for $p \bar p$ 
interactions obtained using the DPD's measured at HERA 
(two upper curves in Fig.~\ref{fig:cdf}).
Breakdown of vertex factorisation between $ep$ and $p \bar p$ interactions
was advocated to explain this result.
The concept of LRG survival probability, 
which accounts for the possibility of secondary emissions which might
fill with particles the LRG in the final state of diffractive events,
was recently revived.
This concept follows and complements 
the line of studies on rescattering corrections
in hadronic interactions (see references in \cite{kaidalov}).
Recently, Kaidalov et al.\cite{kaidalov} 
made a parameter-free computation of  
the LRG survival probability using ISR and Tevatron 
soft scattering data.
The $F_{jj}^D(\beta)$ they obtained
using the LRG survival probability they computed, 
together with the HERA DPD's,
is in surprising agreement with the CDF data
(four bottom curves in Fig.~\ref{fig:cdf}).
The uncertainty in the calculation 
is dominated by the uncertainty in the $\P$ structure functions.
This result supports the universality of DPD's.

\section{Exclusive production of vector mesons at HERA}\label{sec:vm}

The exclusive (or elastic) production of vector mesons (VM) at HERA 
($ep \rightarrow eVp$, where 
$V= \rho^0, \omega, \phi, J/\psi, \psi^\prime$ or $\Upsilon$)
is a very clean experimental process.
As it is measured in a wide kinematic range,
$0<Q^2 < 100 ~{\rm GeV}^2,~20<W
<290~{\rm GeV},~0<|t|<20~{\rm GeV}^2$,
it constitutes an important process to study the dynamics of strong interactions
and allows simultaneous control of the possible hard scales, 
$Q^2, t$ or the squared VM mass $M_{VM}^2$.

\begin{figure}
\epsfxsize180pt
\figurebox{120pt}{190pt}{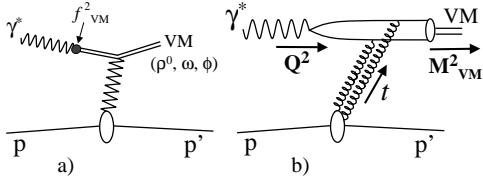}
\caption{Diagrams for elastic VM scattering at HERA, $ep \rightarrow eVp$:
a) soft VM production, b) hard VM production.
}
\label{fig:elastic}
\end{figure}

The elastic photoproduction ($Q^2 \approx 0$) of light VM 
($\rho^0, \omega, \phi$) is measured\cite{crittenden} to be a soft process,
since it shows the properties typical of soft diffraction 
(see Sect.~\ref{subsec:def}).
It can be described by the Vector Dominance Model and the Regge theory.
In this framework, the photon fluctuates into a VM prior to the
interaction, followed by an elastic $Vp \rightarrow Vp$ scatter, 
as shown in Fig.~\ref{fig:elastic}a.
The picture becomes quite different 
if a hard scale is present, and therefore pQCD is expected to be applicable. 
In this case, the elastic VM production can be seen 
as a three-step process (see Fig.~\ref{fig:elastic}b): 
in the target frame, the $\gamma^\star$ fluctuates into a $q \bar q$ dipole,
which then scatters off the proton by a colour-singlet 
two-gluon exchange (at the lowest order), 
and finally the VM is formed, well after the interaction. 
If the dipole size $r=1/\sqrt{z(1-z)Q^2+m_q^2}$ is small,
i.e. we are either in the presence of a large quark mass $m_q$ or
of a longitudinal $\gamma^\star$ of high virtuality,
the $q \bar q$ pair is able to resolve the gluons in the proton, 
and thus pQCD is applicable.
In this case, the cross section is expected to behave in a different way
than in soft diffraction. 
Indeed, pQCD calculations\cite{frankfurt} for longitudinal photons predict:
{\it i)} a fast rise with energy, 
$\sigma_{_L} \propto [1/Q^6] \cdot \alpha_s^2(Q_{eff}^2) \cdot [xg(x,Q_{eff}^2)]^2\propto W^{0.8}$,
since $xg(x,Q^2_{eff})$ is measured at HERA to be $\approx x^{-0.2}$ and 
$x\approx 1/W^2$ at small $x$;
{\it ii)} a tendency to approach the universality of the $t$-dependence, 
namely $d\sigma/dt \propto e^{-b_{2g}|t|}$, with $b_{2g} \sim 4$~GeV$^{-2}$,
almost independent of $W$ 
(i.e. $\alpha_{_\P}^\prime\rightarrow 0 \Rightarrow$~small shrinkage);
{\it iii)} an approximate restoration of flavour independence,
i.e. the $\gamma^\star$ couples directly to the constituent quarks
and the $\rho^0:\omega:\phi:J/\psi$ ratios are expected to
converge to the {\it modified} SU(4) ratios 
$9:1{\bf \cdot 0.8}:2{\bf \cdot 1.2}:8{\bf \cdot 3.5}$
when the hard scale $Q_{eff}^2$ is large enough that 
the VM wave functions converge to an asymptotic form\cite{frankfurt}.

It is important to verify if these predictions are supported by the data. 
It is also interesting to investigate
at which scale the $xg(x,Q_{eff}^2)$ should be evaluated, 
i.e. which or which combination of $Q^2, t$ and $M_{VM}^2$ 
enter in $Q_{eff}^2$.

\subsection{The HERA measurements}

\begin{figure}
\epsfxsize180pt
\figurebox{120pt}{190pt}{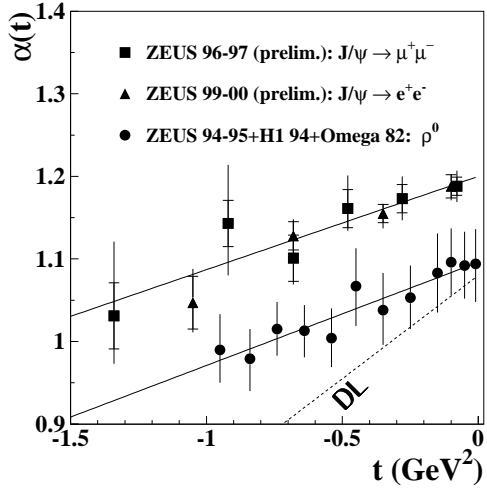}
\caption{Pomeron trajectories extracted from elastic photoproduction
of $\rho^0$ and $J/\psi$ mesons. The dashed line marked DL shows the 
soft $\P$ trajectory\protect\cite{DL}.
}
\label{fig:traj}
\end{figure}

The elastic process $ep \rightarrow eVp$
has been extensively  studied at HERA for many years.
One of the earliest results was the determination in photoproduction of the 
$W^{\sim 0.22}$ dependence\cite{crittenden} 
for light VM's ($\rho^0, \omega$ and $\phi$)
while the $J/\psi$ dependence\cite{crittenden} was more like $W^{0.8}$. 
The behaviour of $\sigma_{\gamma p \rightarrow J/\psi p}$ 
at $Q^2 \approx 0$, is described by the model of Ryskin\cite{ryskin}
when the steep gluon density measured at HERA is used.
This result shows that the $J/\psi$ mass provides a 
hard enough scale to apply pQCD even in photoproduction.
ZEUS measured\cite{rhotraj,jpsitraj} 
the double differential elastic cross sections 
as a function of $W$ and $t$
for $\rho^0, \phi$ and $J/\psi$ photoproduction.
By fitting the $W$ dependence in the $t$ bins, the pomeron 
trajectories 
responsible for the production of the different VM's can be extracted.
The results, shown in Fig.~\ref{fig:traj} and Table~\ref{tab:traject},
show that, when extracted this way, 
the measured $\P$ trajectory is not universal.
In particular, in the case of the $J/\psi$ 
both $\alpha_{_\P}(0)$ and $\alpha_{_\P}^\prime$ 
support the pQCD predictions 
{\it i)} and {\it ii)} of Sect.~\ref{sec:vm}.
\begin{table}
\caption{Pomeron trajectories: the first three entries refer to 
photoproduction and the fourth to DIS.
For comparison,  
the DL soft pomeron\protect\cite{DL} is also listed.}\label{tab:traject}
\begin{tabular}{|c|c|c|} 
\hline 
\raisebox{0pt}[12pt][6pt]{} & 
\raisebox{0pt}[12pt][6pt]{$\alpha_{_\P}(0)$} &
\raisebox{0pt}[12pt][6pt]{$\alpha_{_\P}^\prime $[GeV]$^{-2}$} \\
\hline
\raisebox{0pt}[12pt][6pt]{$\rho^0$} & 
\raisebox{0pt}[12pt][6pt]{$1.096\pm0.021$} & 
\raisebox{0pt}[12pt][6pt]{$0.125\pm0.038$} \\
\raisebox{0pt}[12pt][6pt]{$\phi$} & 
\raisebox{0pt}[12pt][6pt]{$1.081\pm0.010$} & 
\raisebox{0pt}[12pt][6pt]{$0.158\pm0.028$} \\
\raisebox{0pt}[12pt][6pt]{$J/\psi$} & 
\raisebox{0pt}[12pt][6pt]{$1.198\pm0.012$} & 
\raisebox{0pt}[12pt][6pt]{$0.114\pm0.025$} \\
\raisebox{0pt}[12pt][6pt]{$\rho^0 (DIS)$} & 
\raisebox{0pt}[12pt][6pt]{$1.14\pm0.03$} & 
\raisebox{0pt}[12pt][6pt]{$0.04^{+0.15}_{-0.08}$} \\
\raisebox{0pt}[12pt][6pt]{DL soft} & 
\raisebox{0pt}[12pt][6pt]{$1.08$} & 
\raisebox{0pt}[12pt][6pt]{$0.25$}
\\\hline
\end{tabular}
\end{table}

\begin{figure}
\epsfxsize190pt
\figurebox{120pt}{190pt}{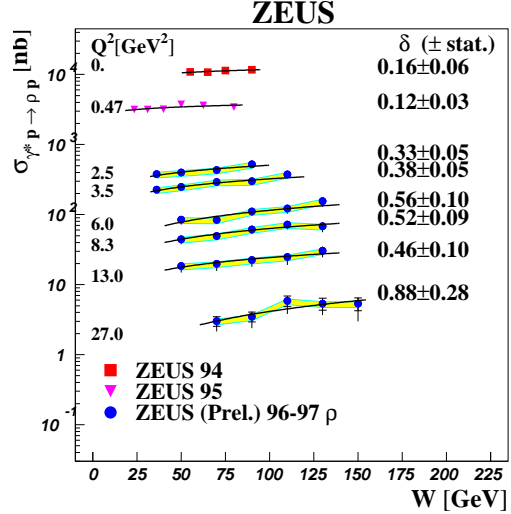}
\caption{Elastic $\rho^0$ cross sections vs. $W$ for several values of $Q^2$.
The results of the fits in the various $Q^2$ bins
with a $W^\delta$ dependence are shown.
}
\label{fig:disrho}
\end{figure}

The ZEUS measurement\cite{arik} of the $ep \rightarrow e\rho^0p$ 
deep-inelastic cross sections
as a function of $W$ in bins of $Q^2$ 
from $Q^2 \approx 0$ to $Q^2 = 27$ GeV$^2$
are shown in Fig.~\ref{fig:disrho}.
A fit to $\sigma_{ep \rightarrow e\rho^0p}$ with a $W^\delta$ dependence
shows that, within the $Q^2$ range measured, $\delta$ varies 
from $\delta = 0.16\pm0.06$ typical of soft scattering
to $\delta = 0.88 \pm 0.22$, close to the value found 
in the case of $J/\psi$ photoproduction.
This transition from the soft to the hard regime proves that 
also large values of $Q^2$ provide
a hard scale to apply pQCD to elastic VM production at HERA.
The result of the extraction of the pomeron trajectory from these data
is also shown in Table~\ref{tab:traject}.
Though statistically limited, there is a hint that 
the pomeron trajectory extracted from $\rho^0$ in DIS is 
closer to the one from  $J/\psi$ in photoproduction than to the soft DL one.\\
An early prediction of pQCD is a different $Q^2$ dependence for
the longitudinal ($\sigma_{_L}$) and transverse ($\sigma_{_T}$)
 cross sections. 
This prediction\cite{martin} was verified
by plotting the ratio $R=\sigma_{_L}/\sigma_{_T}$ 
as a function of $Q^2$;
the result\cite{arik} (not shown) 
is that $R$ is found to increase with $Q^2$ 
as expected in pQCD.

\begin{figure}
\epsfxsize185pt
\figurebox{120pt}{190pt}{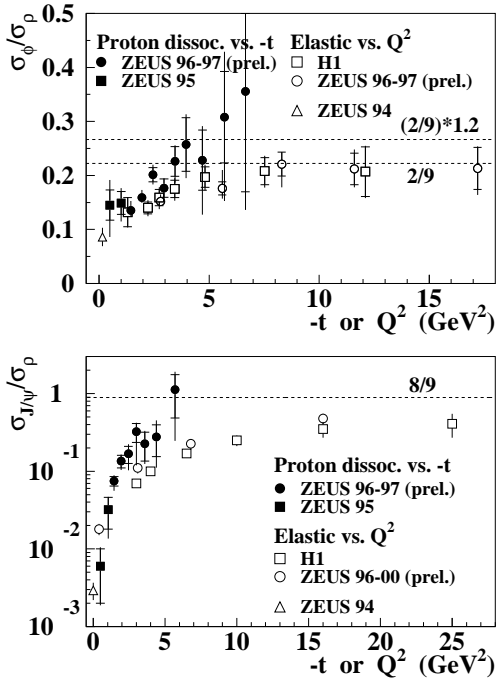}
\caption{Ratio of $\sigma_{\phi}/\sigma_{\rho}$  
and $\sigma_{J/\psi}/\sigma_{\rho}$ cross sections
for elastic production vs. $Q^2$ and 
for proton dissociative production vs. $-t$.
}
\label{fig:vmratio}
\end{figure}

Another interesting quantity to study is 
the ratio of VM production cross sections.
Fig.~\ref{fig:vmratio} shows that the
$\sigma_\phi / \sigma_{\rho}$ and  $\sigma_{J/\psi} / \sigma_{\rho}$
ratios for elastic scattering at HERA
are measured to increase as a function of $Q^2$.
In the case of ${\phi}$ and ${\rho^0}$,
the cross-section ratio approaches the asymptotic value given by the
modified SU(4) factors,
while in the case of ${J/\psi}$ and ${\rho^0}$ the ratio is
still rising at the largest $Q^2$ measured, 
in agreement with pQCD predictions\cite{frankfurt}.
The ZEUS preliminary  ratios 
for the proton dissociative
cross sections $ep \rightarrow eVY$ as a function of $|t|$, 
are also shown in Fig.~\ref{fig:vmratio}. 
Remarkably, the data show that the ratios also rise  
with increasing $|t|$,
which indicates that large values of $|t|$ also
constitute a hard scale,
like $Q^2$ and the VM mass.
The faster rise with  $t$ than with $Q^2$ of both ratios
suggests that $Q^2$ and $t$ might not be equivalent scales, 
in the sense that the cross sections do not depend on them in the same way.

\begin{figure}
\epsfxsize185pt
\figurebox{120pt}{190pt}{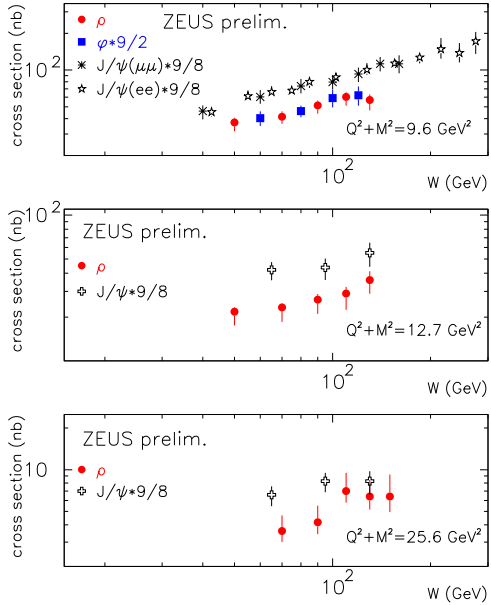}
\caption{Comparison of scaled elastic VM cross sections at fixed
$Q^2+M_{VM}^2$ values. The cross sections are weighted 
as indicated in the figure.
}
\label{fig:aharon}
\end{figure}
It was proposed\cite{h1phi} that the elastic VM cross sections
might show a universal (i.e. independent of VM) behaviour vs.
the variable $Q^2+M_{VM}^2$
after scaling the cross sections by the SU(4) factors $9:1:2:8$.
The new and more precise ZEUS data\cite{bruce}
(Fig.~\ref{fig:aharon}) show 
that, while the scaled $\rho^0, \omega$ and $\phi$ cross sections
lie one on top of the other when plotted vs. $Q^2+M_{VM}^2$,
the scaled $J/\psi$ cross sections are measured to be larger.
The conclusion is that 
the behaviour advocated in~\cite{h1phi} 
works for light VM production
-- when the VM masses are close to each other -- 
but not for $J/\psi$ production.

\section{Selection of recent developments}

\subsection{Deeply virtual Compton scattering
and generalised parton distributions}

The deeply virtual Compton scattering (DVCS) is the exclusive
production of a real photon in DIS, $ep \rightarrow e\gamma p$.
\begin{figure}
\epsfxsize90pt
\figurebox{120pt}{90pt}{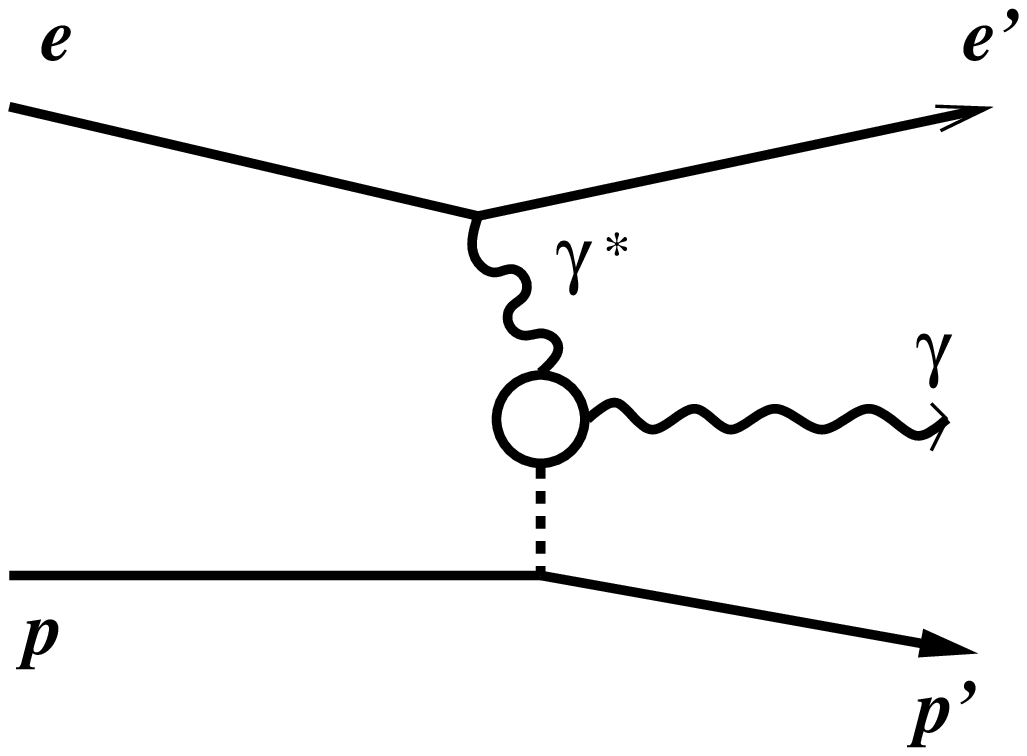}
\epsfxsize100pt
\figurebox{120pt}{100pt}{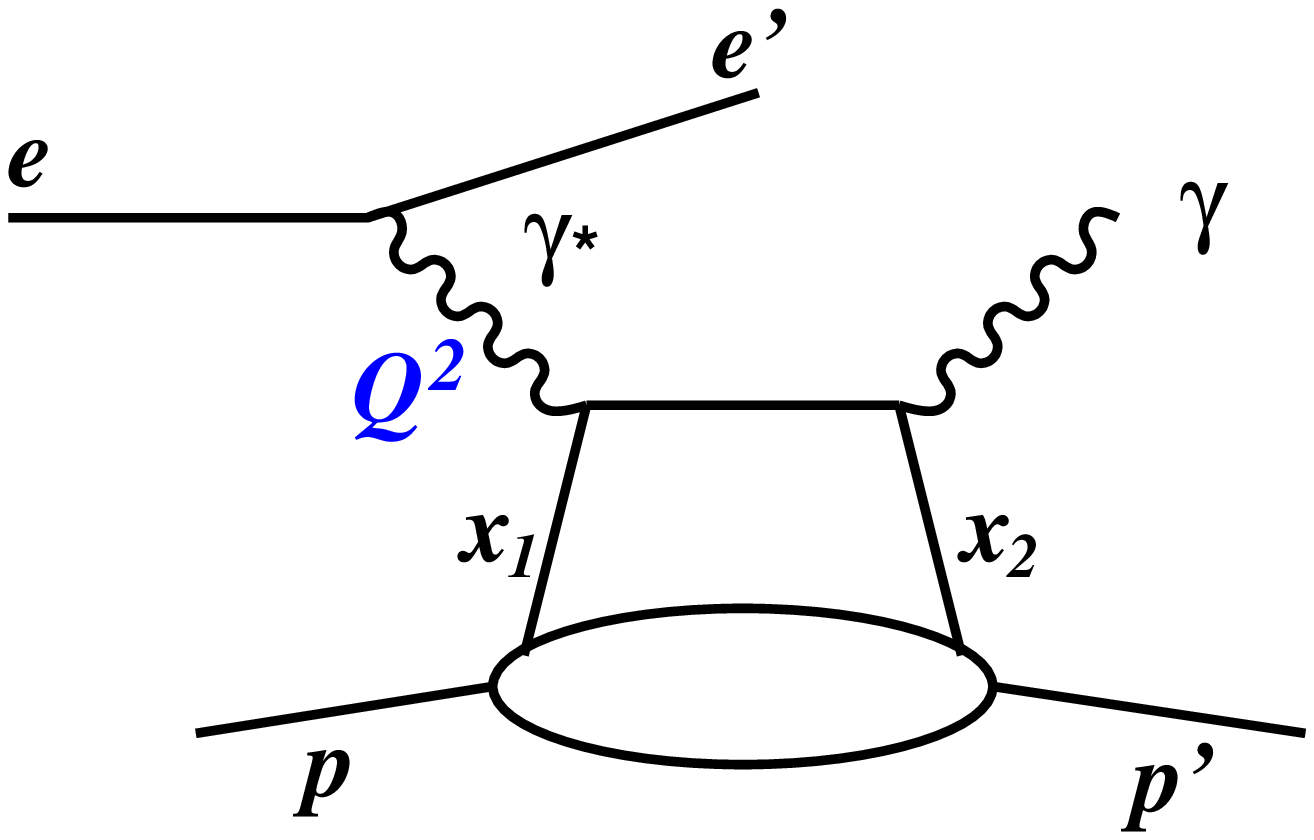}
\caption{Left: generic diagram for DVCS. Right: one of the leading diagrams 
for DVCS in QCD.
}
\label{fig:dvcs}
\end{figure}
This reaction, shown in Fig.~\ref{fig:dvcs}, 
is similar to the elastic VM production, with the difference being
that a real photon appears in the final state instead of a VM.
Therefore the DVCS is theoretically simpler, since no VM wave function 
(a non-perturbative quantity) is involved.
The DVCS process is considered particularly interesting 
since it gives access
to the real part of the scattering amplitude and to the 
generalised parton distributions\cite{gpd} (GPD's). 
The GPD's are fundamental
quantities for exclusive processes in QCD
since they unify the concepts of parton distributions and 
of hadronic form factors.
Indeed, the blob in Fig.~\ref{fig:dvcs}-left can be resolved 
at the lowest order in QCD into a
two-parton exchange (e.g. Fig.~\ref{fig:dvcs}-right) in which, 
to allow for $t = (p - p^\prime)^2 \not= 0$,
the fractional transverse and
longitudinal momenta of the two exchanged partons must be 
different\footnote{The reader should notice that 
the condition $t = (p - p^\prime)^2 \not= 0$
is in general true in diffractive scattering, 
and therefore GPD's should be used. However, most of 
the data shown so far are at small $|t|$, where the GPD's 
can be successfully approximated by the usual parton distributions.},
$k_{t1} \not= k_{t2}$ and $x_1 \not= x_2$.
The usual parton distributions, for which $x_1 = x_2 = x$, 
are obtained 
from the squared wave functions for all partonic configurations containing
a parton with the specified longitudinal momentum $x$,
and therefore represent the probability to find such a parton.
In contrast, the GPD's represent the interference of different wave functions,
and thus correlate different parton configurations in the hadron 
at the quantum-mechanical level\footnote{For this reason the GPD's are also 
called skewed or non-diagonal or off-forward parton distributions.}.
Therefore, in addition to the usual longitudinal momentum $x$, the
GPD's account for parton $k_T$ and two-particle correlations
in the proton.

\begin{figure}
\epsfxsize190pt
\figurebox{120pt}{190pt}{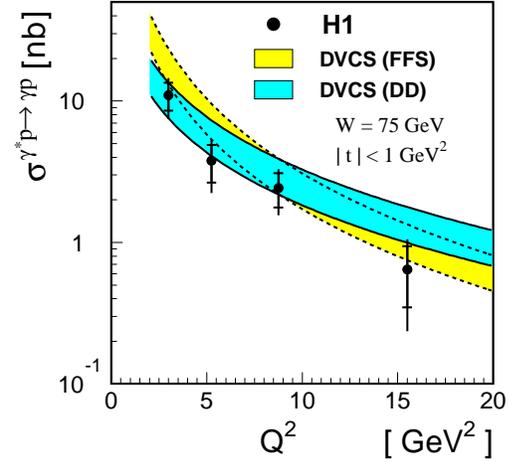}
\caption{The DVCS cross section vs. $Q^2$
compared to QCD-based calculations\protect\cite{dvcspred1,dvcspred2}.
}
\label{fig:h1dvcs}
\end{figure}

The DVCS process was searched for by 
H1\cite{h1dvcs} and ZEUS\cite{zeusdvcs}.
An excess of photons over the expectations of the QED-Compton
background process was observed and the  
cross section for the DVCS process was measured 
as a function of $W$ and $Q^2$.
The results (see Fig.~\ref{fig:h1dvcs}) 
are in agreement with the QCD-based predictions\cite{dvcspred1,dvcspred2}.
 
\subsection{Generalised parton distributions from inclusive 
diffraction at large $\beta$}

\begin{figure}
\epsfxsize190pt
\figurebox{120pt}{190pt}{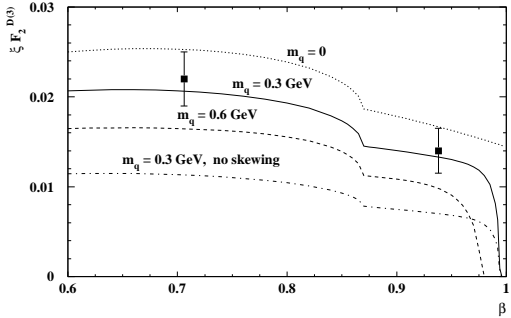}
\caption{$F_2^D$ at $Q^2 = 60$ GeV$^2$ and $x_{_\P}=0.0042$ 
for three different choices of the light quark mass as indicated 
in the plot. The ZEUS data\protect\cite{mx2} (squares) are also shown. 
The dash-dotted line is the prediction if the effects 
from skewing are neglected.
}
\label{fig:hebecker}
\end{figure}

Hebecker and Teubner suggested\cite{hebecker} that the 
generalised gluon distribution
can be extracted from inclusive diffractive electroproduction  
at large $Q^2$ and $\beta\rightarrow 1$.
They demonstrated that, within certain restrictions in the kinematic domain,
the process is, in principle, perturbatively calculable and highly 
sensitive to effects due to the non-diagonal parton distributions
(skewing effects). 
A leading order numerical analysis, 
which includes corrections for the skewdness of the parton distributions 
and for the real part of the amplitude,  
and an estimate of the NLO effects,  
is consistent with the $F_2^D$ data (see Fig.~\ref{fig:hebecker}).
The results  show
the strong sensitivity of $F_2^D$ to skewing effects,
which amount to approximately a factor of two.
Reversing the argument, they assert that
precise data at higher $Q^2$ and full NLO calculations 
should allow the extraction of GPD's with high accuracy.

\subsection{Search for the odderon}

In the framework of the Regge theory, it is possible to introduce
a new trajectory, the 
C = P = --1
partner of the pomeron trajectory, 
which would contribute with different signs to the
particle-particle and particle-antiparticle total cross sections\cite{odderon}.
The exchange of this C-odd trajectory would 
therefore be responsible for any difference at asymptotic energies
between these two total cross sections.
For this reason, this hypothetical exchange was  called 
odderon trajectory, from ``odd-under-exchange''.
No experimental sign of  odderon exchange has been found 
in hadronic scattering. \\
QCD predicts the existence of the odderon,
which can be represented at the lowest order by 
the exchange of three gluons, whereas the pomeron corresponds to the 
exchange of two gluons.
Both perturbative QCD predictions\cite{vacca} for the process
$\gamma p \rightarrow \eta_c p$ 
and non-perturbative QCD predictions\cite{qcdodd} 
(based on the stochastic vacuum model)
for the processes $\gamma p \rightarrow \pi^0 X$ and
$\gamma p \rightarrow f_2(1270) X$ 
at HERA are available.

H1 made a search for odderon exchange 
by studying multi-photon final states
in diffractive events. 
This is an experimentally-clean QCD test 
with purely electromagnetic final states, except for the proton.
Since the C-parity of the exchange fixes the
number of photons in the final state, 
pomeron-mediated diffractive processes at HERA
-- such as $\gamma p \rightarrow \omega p$ and 
$\gamma p \rightarrow \omega \pi^0 X$ -- have an even number 
of photons in the final state,
while odderon-mediated diffractive processes -- such as
$\gamma p \rightarrow \pi^0 X$,
$\gamma p \rightarrow f_2(1270) X$ and
$\gamma p \rightarrow a^0_2(1320) X$ --
would manifest themselves by the presence of 
an odd number of photons in the final state.
The result\cite{h1odderon} is that, while signals are found for the
pomeron-mediated processes and the measured cross sections are 
in agreement with previous measurements in other decay channels,
no signal is found in any of the three odderon-mediated processes 
mentioned above.

\section{Conclusions}

It is not an easy task to draw conclusions on such a complicated
topic such as diffractive scattering.  
For sure, the multitude of hard-diffraction studies at hadron colliders 
and at HERA and the huge theoretical effort 
to describe these phenomena in the framework of perturbative QCD,
contributed to bridge 
between the hadronic  
--large distance-- degrees of freedom typical of Regge theory 
and the partonic degrees of freedom
which unveil the dynamics of strong interactions.
An important piece is still missing, which should allow the description of
large-distance processes within the framework of quantum-field theory.
This is the field in which the study of diffractive
phenomena is expected to provide useful data.
Again, the problem turns
into the understanding of the dynamics of
colour radiation at large-distances, which is a long standing problem
in QCD, that eventually coincides with the understanding of 
confinement of hadrons, one of the few remaining puzzles of  
the Standard Model of particle physics.

\section*{Acknowledgments}
I thank J. Bartels, T. Carli and J. Whitmore for 
their comments on an earlier version of this manuscript.
I profited by the discussions with
J. Bartels, ${\rm W.~Buchm\ddot uller}$, M. Diehl, K. Golec-Biernat, G.P. Vacca
and by the forbearance of the ZEUS and H1 colleagues.
I wish to express my gratitude 
to the DESY directorate for partial support and for making DESY
a challenging and sympathetic research environment. 

\noindent 
{\bf Discussion}

\noindent
B. Ward, MPI Munich and Univ. of Tennessee:
You showed that the model of Golec-Biernat and 
${\rm W\ddot u sthoff}$ fits the data better without DGLAP evolution, 
but you did not say what conclusion you draw from this. 
What is your conclusion? \\
Answer:
In the original version of the model, 
the dipole cross section, which contains the dynamics, was postulated
without any assumption on the evolution, 
and the three free parameters of the model were determined by a fit
 to inclusive DIS data.
Therefore, the agreement of the model with diffractive data,
obtained using those parameters, can be
regarded as a prediction.
The fact that the model modified 
to include DGLAP evolution 
gives a worse description of the diffractive data if the
same parameters are used, might indicate 
either that those parameters are no more adequate
or that terms  beyond the DGLAP equation are needed in the evolution.



\end{document}